\documentclass[12pt]{article}
\usepackage{amsfonts}
\usepackage{amssymb}

\def\hybrid{\topmargin -20pt    \oddsidemargin 0pt
        \headheight 0pt \headsep 0pt
        \textwidth 6.35in       
        \textheight 9.25in       
        \marginparwidth .875in
        \parskip 5pt plus 1pt   \jot = 1.5ex}

\hybrid

\def\baselinestretch{1.2}

\catcode`\@=11

\def\marginnote#1{}
\newcount\hour
\newcount\minute
\newtoks\amorpm
\hour=\time\divide\hour by60
\minute=\time{\multiply\hour by60 \global\advance\minute by-\hour}
\edef\standardtime{{\ifnum\hour<12 \global\amorpm={am}%
        \else\global\amorpm={pm}\advance\hour by-12 \fi
        \ifnum\hour=0 \hour=12 \fi
        \number\hour:\ifnum\minute<10 0\fi\number\minute\the\amorpm}}
\edef\militarytime{\number\hour:\ifnum\minute<10 0\fi\number\minute}

\def\draftlabel#1{{\@bsphack\if@filesw {\let\thepage\relax
   \xdef\@gtempa{\write\@auxout{\string
      \newlabel{#1}{{\@currentlabel}{\thepage}}}}}\@gtempa
   \if@nobreak \ifvmode\nobreak\fi\fi\fi\@esphack}
        \gdef\@eqnlabel{#1}}
\def\@eqnlabel{}
\def\@vacuum{}
\def\draftmarginnote#1{\marginpar{\raggedright\scriptsize\tt#1}}

\def\draft{\oddsidemargin -.5truein
        \def\@oddfoot{\sl preliminary draft \hfil
        \rm\thepage\hfil\sl\today\quad\militarytime}
        \let\@evenfoot\@oddfoot \overfullrule 3pt
        \let\label=\draftlabel
        \let\marginnote=\draftmarginnote
   \def\@eqnnum{(\theequation)\rlap{\kern\marginparsep\tt\@eqnlabel}%
\global\let\@eqnlabel\@vacuum}  }

\def\preprint{\twocolumn\sloppy\flushbottom\parindent 2em
        \leftmargini 2em\leftmarginv .5em\leftmarginvi .5em
        \oddsidemargin -.5in    \evensidemargin -.5in
        \columnsep .4in \footheight 0pt
        \textwidth 10.in        \topmargin  -.4in
        \headheight 12pt \topskip .4in
        \textheight 6.9in \footskip 0pt
        \def\@oddhead{\thepage\hfil\addtocounter{page}{1}\thepage}
        \let\@evenhead\@oddhead \def\@oddfoot{} \def\@evenfoot{} }

\def\numberbysection{\@addtoreset{equation}{section}
        \def\theequation{\thesection.\arabic{equation}}}

\def\underline#1{\relax\ifmmode\@@underline#1\else
        $\@@underline{\hbox{#1}}$\relax\fi}

\def\titlepage{\@restonecolfalse\if@twocolumn\@restonecoltrue\onecolumn
     \else \newpage \fi \thispagestyle{empty}\c@page\z@
        \def\thefootnote{\fnsymbol{footnote}} }

\def\endtitlepage{\if@restonecol\twocolumn \else \newpage \fi
        \def\thefootnote{\arabic{footnote}}
        \setcounter{footnote}{0}}  

\catcode`@=12
\relax

\def\figcap{\section*{Figure Captions\markboth
        {FIGURECAPTIONS}{FIGURECAPTIONS}}\list
        {Figure \arabic{enumi}:\hfill}{\settowidth\labelwidth{Figure
999:}
        \leftmargin\labelwidth
        \advance\leftmargin\labelsep\usecounter{enumi}}}
 \relax
\def\tablecap{\section*{Table Captions\markboth
        {TABLECAPTIONS}{TABLECAPTIONS}}\list
        {Table \arabic{enumi}:\hfill}{\settowidth\labelwidth{Table
999:}
        \leftmargin\labelwidth
        \advance\leftmargin\labelsep\usecounter{enumi}}}
 \relax
\def\reflist{\section*{References\markboth
        {REFLIST}{REFLIST}}\list
        {[\arabic{enumi}]\hfill}{\settowidth\labelwidth{[999]}
        \leftmargin\labelwidth
        \advance\leftmargin\labelsep\usecounter{enumi}}}
 \relax
\makeatletter
\newcounter{pubctr}
\def\publist{\@ifnextchar[{\@publist}{\@@publist}}
\def\@publist[#1]{\list
        {[\arabic{pubctr}]\hfill}{\settowidth\labelwidth{[999]}
        \leftmargin\labelwidth
        \advance\leftmargin\labelsep
        \@nmbrlisttrue\def\@listctr{pubctr}
        \setcounter{pubctr}{#1}\addtocounter{pubctr}{-1}}}
\def\@@publist{\list
        {[\arabic{pubctr}]\hfill}{\settowidth\labelwidth{[999]}
        \leftmargin\labelwidth
        \advance\leftmargin\labelsep
        \@nmbrlisttrue\def\@listctr{pubctr}}}
 \relax
\makeatother
\newskip\humongous \humongous=0pt plus 1000pt minus 1000pt

\newif\ifdtup

\relax

\def\be{\begin{equation}}
\def\ee{\end{equation}}
\def\ba{\begin{eqnarray}}
\def\ea{\end{eqnarray}}

\def\del{\partial}

\def\r{\rho}
\def\a{\alpha}

\def\b{\beta}

\def\g{\gamma}

\def\d{\delta}

\def\e{\epsilon}

\def\th{\theta}

\def\om{\omega}

\def\s{\sigma}

\def\no{\noindent}

\def\qq{\qquad}

\def\IR{\relax{\rm I\kern-.18em R}}

\def \ha {{1\over 2}}

\def \ov {\over}

\def\IR{\relax{\rm I\kern-.18em R}}
\def\inv{^{\raise.15ex\hbox{${\scriptscriptstyle -}$}\kern-.05em 1}}

\begin{document}

\newcommand{\beq}{\begin{equation}}
\newcommand{\eeq}[1]{\label{#1}\end{equation}}
\newcommand{\ber}{\begin{eqnarray}}
\newcommand{\eer}[1]{\label{#1}\end{eqnarray}}
\newcommand{\eqn}[1]{(\ref{#1})}
\begin{titlepage}
\begin{center}

\hfill hep--th/0206091\\
\hfill June 2002\\
\vskip .6in

{\Large \bf String backgrounds and LCFT}

\vskip 0.6in

 {\bf Konstadinos Sfetsos}
\vskip 0.1in

\vskip .1in

Department of Engineering Sciences, University of Patras\\
26110 Patras, Greece\\
{\footnotesize{\tt sfetsos@mail.cern.ch, des.upatras.gr}}\\

\end{center}

\vskip .6in

\centerline{\bf Abstract}
\no
We describe a large class of exact string backgrounds
with a null Killing vector arising,
via a limiting \`a la Penrose procedure, from string
backgrounds corresponding to coset conformal field theories for compact
groups $G_N/H_N$
times a free time-like boson $U(1)_{-N}$.
In this way a class of novel logarithmic conformal field theories (LCFT)
emerges, that includes the one constructed recently as an $N\to \infty$ limit
of the $SU(2)_N/U(1) \times U(1)_{-N}$ theory.
We explicitly give the exact operator algebra for the basic chiral fields
as well as their representation in terms of free bosons,
even though these are not known in general at finite $N$. We also compute
four-point functions of various operators in the theory.
For the cases of the four- and five-dimensional models, corresponding to a
limit of the theory $SO(D+1)_N/SO(D)_N \times U(1)_{-N}$ for $D=3$ and 4,
we also present the explicit expressions for the background fields.

\noindent

\vskip .4in
\noindent

\end{titlepage}
\vfill
\eject

\def\baselinestretch{1.2}

\baselineskip 20pt

\section{Introduction}

It was recently realized that novel logarithmic conformal field theories
arise in the large level limit of coset conformal field
theories for compact groups combined with a free time-like boson \cite{BSpp}.
The example considered in \cite{BSpp} corresponds to the
$SU(2)_N/U(1)_N \times U(1)_{-N}$ theory at $N\to \infty$ (taken in a
correlated way) and has the space-time interpretation
of a three-dimensional plane wave. At the level of
the string background fields, this procedure
corresponds to a Penrose-type limit \cite{Pen}.
At the level of the chiral algebra involving
the compact parafermions \cite{paraf}
and the $U(1)$-current corresponding to the
free time-like boson, one forms linear combinations of the fields which
have well defined operator algebra when $N\to \infty$ and as it turns
out they define a logarithmic
conformal field theory.\footnote{These theories originated in the work
of \cite{saleur,Gur} and have been the subject of
intense research over the past few years
for various diverse reasons \cite{BK}-\cite{persis},
starting with condensed matter physics \cite{gardy,saleur2}
(for a recent review see \cite{Flrev} and references therein).}
This limiting
procedure\footnote{We note that it is
similar to what is called Saletan contraction
for Lie algebras (see, for instance, \cite{Gilmore}) that differs from
the more popular Inonu--Wigner contraction which
in our case would have led
to flat space-times and trivial conformal field theories.} has already
appeared in string theory some years ago where was
used to construct plane wave solutions starting from WZW model
current algebra theories or gauged WZW model coset theories
\cite{sfe1}-\cite{sfetse}.
The same results can also be obtained by working directly
with non-semisimple groups \cite{NaWi}-\cite{Antoniadis}, which however,
as it turned out in all known cases, correspond to contractions of direct
products groups.

The essential reason that a logarithmic structure for the conformal field
theory arose in \cite{BSpp} is that the original
$SU(2)_N/U(1)_N \times U(1)_{-N}$ theory
contains as basic chiral operators the lowest parafermions with
conformal dimension $1-1/N$ as well as
the $U(1)$ current, corresponding to the free time-like boson,
with dimension exactly $1$ for all $N$.
Hence, although classically both operators have dimension $1$,
quantum mechanically this dimension is protected only in the case of the
$U(1)$ current. The remnant of this, in the limit $N\to \infty$,
is the fact that
among the three basic fields that emerge, two are Virasoro primaries, but one
of the them has the third field as, its so-called, logarithmic partner, which
is not a Virasoro primary. It is worth here to make the distinction that,
such logarithmic conformal field theories did not occur in the similar
contraction of current algebra theories \cite{Olive,sfe2}
since the combination of
currents that one forms in order to take the limit $N\to \infty$,
has well defined conformal dimension $1$ for all $N$.

The purpose of the present paper is to investigate general higher dimensional
theories corresponding to
coset models for compact groups $G_N/H_N \times U(1)_{-N}$ in the limit
$N\to \infty$. Unlike the three-dimensional example of \cite{BSpp} we do not
have a well studied theory for the non-abelian parafermions that also form
the basic chiral operators in this case and we do not know their operator
product algebra unambiguously. However, guided by their well known
classical Poisson algebra constructed in \cite{claspara}
and the insight gained from the example of
\cite{BSpp}, we will develop a free field representation in terms of bosons.
We will also explicitly construct the operator product algebra and compute
the four-point functions of the basic operators in the theory.
We emphasize that this is done without the prior knowledge of these at
finite $N$. We will find out that the space-time
metric of the string background is not in general a plane wave, but it belongs
to the more general class of metrics with a null Killing vector
(for various aspects concerning such backgrounds, see
\cite{kill1}-\cite{kill2}) in which
plane waves form a particular subclass.
This will be demonstrated explicitly for the
four- and five- dimensional string
backgrounds corresponding to the limits of the
$SO(D+1)_N/SO(D)_N\times U(1)_{-N}$ theories for $D=3$ and $4$.

The organization of this paper is as follows: In section 2 we define our models
and treat them at the semi-classical level. In section 3 we develop
generally and systematically the free field representation of the various
operators at the quantum level and also compute their four-point functions.
In section 4 we present the explicit string backgrounds for some
four- and five-dimensional
examples and finally end the paper in section 5 with concluding remarks.

\section{General models; semi-classical treatment}

We start with the coset conformal field theory model $G_N/H_N$ at level $N$
for a general compact group $G$ and its subgroup $H$. The Lagrangian
description of such models is given in terms of the corresponding
gauged WZW model action \cite{karabali}.
The basic chiral operators in this theory are the parafermions
which we will denote by $\psi_\a$, where the index
$\a$ takes values over the coset space $G/H$, i.e. $\a=1,2,\dots ,\dim(G/H)$.
At the
classical level in the $1/N$ expansion they obey a Poisson bracket algebra
computed in \cite{claspara}. In a real basis for the parafermions,
it reads (the world-sheet light-cone variable $\s_-$ as the ``time'')
\ba
\{ \psi_\a,\psi_\b\} & =&  -{1\ov \pi} \d_{\a\b} \d^\prime(\s_+-\s'_+) -
{1\ov \sqrt{N}} f_{\a\b\g} \psi_\g(\s'_+) \d(\s_+-\s'_+)
\nonumber\\
&&  -{\pi\ov 2 N} f_{c\a\g} f_{c\b\d}
\e(\s_+-\s'_+) \psi_\g(\s_+)\psi_\d(\s'_+)\ ,
\label{pppor}
\ea
where early latin indices take values over the
subgroup $H$, i.e. $\a=1,2,\dots , \dim(H)$.
We have assumed in writing Poisson brackets that the first parafermion
is evaluated at $\s_+$ and the second at $\s'_+$, whereas the $\s_-$
dependence is common in both of them.
We have also denoted the antisymmetric step function by $\e(x)=1 $ $(-1)$
for $x>0$ $(x<0)$ and by $\d(x)$ and $\d'(x)$ the usual $\d$-function and
its derivative.
In the limit $N\to \infty$ the parafermions become
free bosons as one can easily see.
Nevertheless we will see that by a taking a correlated large
$N\to\infty$ limit involving the $U(1)$ current for the extra time-like boson
we will be left with a non-trivial algebra.
We will denote by $J_0$ the generator corresponding to the extra
time-like $U(1)_{-N}$ factor with Poisson bracket
\be
\{J_0,J_0\}= {N\ov \pi}\d^\prime(\s_+-\s'_+)\ .
\ee
For our purposes we separate one of the parafermions,
say $\psi_0$ and split the coset index
as $\a=(0,i)$, where $i=1,2,\dots \dim(G/H)-1$.
The non-vanishing structure constants in this basis can be
\be
f_{ijk}\ ,\quad f_{cij}\ ,\quad f_{ij0}=f_{ij}\ ,\quad f_{ci}=f_{ci0}\ .
\label{sttru}
\ee
Then we charge basis for the operators as
\be
\Phi = {\a\ov N} (\sqrt{N} \psi_0-J_0)\ ,\quad \Psi = \sqrt{N} \psi_0+J_0\ ,
\quad P_i=\sqrt{\a}\psi_i\ ,\quad i=1,2,\dots , \dim(G/H)-1\ ,
\label{reff}
\ee
where $\a$ is a free parameter (not to be confused with the coset theory
index) and take the limit $N\to \infty$. This is singular
in the sense that the above change of basis is not invertible.
However, as in \cite{BSpp} we find
that the contracted Poisson-bracket algebra for the
$\dim(G)+1$ generators $(\Phi,\Psi,P_i)$ is well defined in that
limit and reads
\ba
\{\Phi,\Phi\} & = &0\ ,
\nonumber\\
\{\Phi,\Psi\}& =& -{2\a\ov \pi}\d'(\s_+-\s'_+)\ ,
\nonumber\\
\{\Psi,\Psi\} & = &
-{\pi\ov 2 \a} f_{ci} f_{cj} P_i(\s_+)P_j(\s'_+)\e(\s_+-\s'_+)\ ,
\nonumber\\
\{P_i,P_j\} & = &
 -{\a\ov \pi}\d_{ij} \d'(\s_+-\s'_+) -\ha f_{ij} \Phi(\s'_+) \d(\s_+-\s'_+)
\nonumber\\
&& -{\pi \ov 8 \a} f_{ci} f_{cj} \Phi(\s_+) \Phi(\s'_+)\e(\s_+-\s'_+)\ ,
\label{pppoi1}\\
\{\Phi,P_i\} & = & 0 \ ,
\nonumber\\
\{\Psi,P_i\} & = & -f_{ij} P_j(\s'_+) \d(\s_+-\s'_+) + {\pi\ov 4 \a}
f_{ci} f_{cj} P_j(\s_+) \Phi(\s'_+)\e(\s_+-\s'_+)\ .
\nonumber\
\ea
Notice that among the non-vanishing structure constants in
\eqn{sttru}, $f_{ijk}$ and $f_{cij}$ do not explicitly appear in \eqn{pppoi1}.
We also note the form of the algebra in the particular case of the model
$SO(D+1)_{N}/SO(D)_N\times U(1)_{-N}$. Since the coset factor
is a symmetric space
we have that $f_{ij}=f_{ijk}=0$ and also for this particular coset case
according to our normalizations
$f_{ci}f_{cj}=\d_{ij}$. Hence \eqn{pppoi1} becomes
\ba
\{\Phi,\Phi\} & = &0\ ,
\nonumber\\
\{\Phi,\Psi\}& =& -{2\a\ov \pi}\d'(\s_+-\s'_+)\ ,
\nonumber\\
\{\Psi,\Psi\} & = &
-{\pi\ov 2 \a} P_i(\s_+)P_i(\s'_+)\e(\s_+-\s'_+)\ ,
\nonumber\\
\{P_i,P_j\} & = &
 -{\a\ov \pi}\d_{ij} \d'(\s_+-\s'_+)
-{\pi \ov 8 \a} \d_{ij} \Phi(\s_+) \Phi(\s'_+)\e(\s_+-\s'_+)\ ,
\label{pppoi2}\\
\{\Phi,P_i\} & = & 0 \ ,
\nonumber\\
\{\Psi,P_i\} & = & {\pi\ov 4 \a} P_i(\s_+) \Phi(\s'_+)\e(\s_+-\s'_+)\ .
\nonumber
\ea
We emphasize that, unlike the parafermionic Poisson bracket
algebra \eqn{pppor}, the algebra \eqn{pppoi1} (and obviously \eqn{pppoi2})
is exact since the parameter $\a$ can be set to
any value by appropriate rescalings of the generators. Therefore without
loss of generality we choose to set $\a=1$ in the rest of the paper.
In order to construct the
operator product expansions for the basic operators in these models
it will be useful to
construct first a representation for them in terms of free bosons.
A priori it is not obvious how this can found since for general groups $G$ and
$H$ there is no such free field representation for the parafermions of
the coset model theory
$G_N/H_N$ at finite $N$. However, one can use the fact that for the case of
the coset $SU(2)_N/U(1)_N$ such a free field representation is well known.

\section{Operator algebra, free fields and correlators}

The free field realization of the theory
obtained from the correlated $N\to \infty$ limit on the
$SU(2)_N/U(1)_N\times U(1)_{-N}$ theory has been given in \cite{BSpp}.
We repeat its derivation here by including some details necessary
for our present purposes.
In this case we have three operators
obeying the algebra \cite{BSpp} (we focus
to the holomorphic part of the theory)
\ba
\Psi(z)\Phi(w) & =&  {1\ov (z-w)^2} + {\cal O}(1) \ ,
\nonumber\\
\Psi(z)\Psi(w)& = & {\ln(z-w)\ov (z-w)^2} + 2 \ln(z-w) :\!P^2(w)\!:
+ \ha \ln^2(z-w) :\!\Phi^2(w)\!: +\ {\cal O}(1)\ ,
\nonumber\\
\Psi(z) P(w)& = &  - \ln(z-w) :\!(P\Phi)(w)\!: +\ {\cal O}(1) \ ,
\label{oppe2}\\
P(z)P(w) & = & {1\ov 2 (z-w)^2} + \ha \ln(z-w) :\!\Phi^2(w)\!:
+\ {\cal O}(1) \ .
\nonumber
\ea
We also note that the fields $\Phi$ and $P$ are primaries of the Virasoro
stress energy tensor of the theory with conformal dimensions one, whereas
the field $\Psi$ is the logarithmic partner of $\Phi$ and obeys
the operator product
\be
T(z)\Psi(w)= {\Psi(w)-\Phi(w)/2\ov (z-w)^2} + {\del \Psi(w)\ov z-w}
+ {\cal O}(1)\ ,
\label{loogg}
\ee
which is a generic characteristic of logarithmic conformal field theories
(see, for instance, \cite{Flrev}).
In order to reproduce this operator algebra
we start from the free field representation of the
coset model $SU(2)_N/U(1)_N$ at finite $N$ which is known \cite{parfree}.
Introducing two real free bosons,
\be
\langle \phi_i(z) \phi_j(w)\rangle = -\delta_{ij} \ln(z-w)\ ,\qq
i, j = 1, 2~,
\ee
one may represent the elementary parafermion currents as\footnote{All
expressions appearing in the rest of this paper to involve products of free
bosons and their derivatives, are understood as being properly normal ordered.}
\ba
\psi_1 &=& {1 \over \sqrt{2}} \left( -\sqrt{1 + {2/N}}\ \partial \phi_1
+ i \partial \phi_2 \right) e^{+\sqrt{2/N}\ \phi_2}\ ,
\nonumber\\
\psi_1^\dagger &=& {1 \over \sqrt{2}} \left( \sqrt{1 + {2/N}}\ \partial \phi_1
+ i \partial \phi_2 \right) e^{-\sqrt{2/N}\ \phi_2}\ .
\label{free1}
\ea
We will denote the time-like free boson for the $U(1)_{-N}$ factor
by $\phi_0$. It obeys
\be
\langle \phi_0(z) \phi_0(w)\rangle = \ln(z-w)\ ,
\ee
and the current itself is given by
\be
J_0=i\sqrt{N\ov 2} \del \phi_0\ .
\label{free2}
\ee
Note that in order for the parafermions
$\psi_1$ and $\psi^\dagger_1 $ to be complex conjugate fields of
each other and the current $J_0$ to be hermitian, the bosons
$\phi_0$, $\phi_1$ and $\phi_2$ are antihermitian. Hence, unlike the previous
section where we used
a real basis for the classical parafermions, here we use a complex basis.
The stress-energy tensor of the entire theory is
\be
T(z)= \ha (\del\phi_0)^2 -\ha (\del\phi_1)^2-\ha (\del\phi_2)^2
-{i\ov \sqrt{2(N+2)}} \del^2 \phi_1 \
\label{strq}
\ee
and corresponds to a central charge $c=3N/(N+2)$.
Let us now consider the scalar field redefinition
\be
\phi_+ = {1\ov \sqrt{2N}} (\phi_0+\phi_1)\ ,\qq
\phi_- = \sqrt{N\ov 2} (\phi_0-\phi_1)\ ,\qq \phi_2= \phi\ ,
\label{coommm}
\ee
Then, the new set of scalars obey
\be
\langle \phi_+(z) \phi_-(w)\rangle = -\langle \phi(z) \phi(w)\rangle =
\ln(z-w)\
\ee
and have zero correlators otherwise.
Then we make the field redefinitions
\be
\Phi={i\ov 2 \sqrt{N}}(\psi_1-\psi_1^\dagger)-{1\ov N}J_0\ ,\qq
\Psi={i\ov 2} \sqrt{N}(\psi_1-\psi_1^\dagger)+J_0\ ,\qq
P=\ha (\psi_1+\psi_1^\dagger)\ ,
\label{redd}
\ee
which mix the parafermions with
the $U(1)$ current.\footnote{Compared to \eqn{reff} one notices that the real
parafermions $\psi_0$ and $\psi_1$ are related to the complex conjugate
ones $\psi_1$ and
$\psi_1^\dagger$ of this section as $\psi_0\to {i\ov 2}(\psi_1-\psi_1^\dagger)$
and $\psi_1\to {1\ov 2}(\psi_1+\psi_1^\dagger)$.}
Then in the limit $N\to \infty$ we find the expansion
\be
\psi_1= -{\sqrt{N}\ov 2} \del\phi_+
+ {1\ov \sqrt{2}} (i \del\phi-\phi \del \phi_+) +{1\ov \sqrt{N}}\left(
{1\ov 2}\del\phi_- + i \phi\del\phi -\ha (\phi^2 + 1) \del\phi_+\right)
+ {\cal O}\left(1\ov N\right)\ ,
\ee
whereas the similar expression for $\psi^\dagger_1$ can be found by
complex conjugation. Using \eqn{redd} we find that in the limit
$N\to \infty$
\ba
&& \Psi=-{i\ov 2} (\phi^2 + 1)\del\phi_+ - \phi \del \phi
+ i \del\phi_-\ ,\qq \Phi= -i \del\phi_+\ ,
\nonumber\\
&& P= {1\ov \sqrt{2}}(i \del\phi - \phi \del\phi_+)\ .
\label{JHWE}
\ea
We can now verify that these obey the operator product expansions in
\eqn{oppe2}.
Also, the stress-energy tensor \eqn{strq} of the theory becomes in this limit
\be
T(z)= \del \phi_+\del\phi_- - \ha (\del \phi)^2
-{i\ov 2}\del^2 \phi_+
\label{ferk}
\ee
and corresponds to a central charge $c=3$ theory.
This form of the energy momentum tensor can also be obtained by demanding
that the operators $\Phi,\Psi$ and $P$, in the free field representation
\eqn{JHWE}, have the correct operator product expansion with $T(z)$.
In particular, notice the presence of
the background charge along the null direction.\footnote{Another way to
derive \eqn{ferk} is to note that the stress energy tensor of the theory
takes the form
\be
T(z)= \Psi \Phi -\ha \Phi^2 + P^2\ ,
\ee
which can be proven using the parafermion algebra.
Substituting the expressions \eqn{JHWE} we prove \eqn{ferk}. The background
charge term arises by taking into account the effect of normal ordering.}
This term does not contribute
to the central charge of the theory, but it is crucial in
verifying that in the representation \eqn{JHWE} the field $\Psi$ is the
logarithmic partner of $\Phi$ and obeys \eqn{loogg}.
We also note for completeness that, had we
interchanged in \eqn{coommm} the bosonic fields $\phi_1$ and $\phi_2$ would not
have led to a consistent limit for the free field theory representation of
\eqn{free1} and \eqn{free2}. In that sense we consider
the free field representation \eqn{JHWE} as rather unique, up to local field
redefinitions.\footnote{For
completeness we also note here that the free field representation for the
current algebra corresponding to the four-dimensional plane wave of
\cite{NaWi}, which was constructed in \cite{KiKouLu}, can also be obtained
as a limit of the free field representation of the $SU(2)_N \times U(1)_{-N}$
current algebra.}

Next we give the free field representation for the case of the model obtained
in the $N\to \infty$ limit of the $SO(D+1)_N/SO(D)_N\times U(1)_{-N}$ theory.
We introduce $D+1$ free bosons denoted by $\phi_+$, $\phi_-$ and $\phi_i$,
$i=1,2,\dots , D-1$ which obey
\be
\langle \phi_+(z) \phi_-(w)\rangle = \ln(z-w)\ ,\qq
\langle \phi(z) \phi(w)\rangle =-\d_{ij} \ln(z-w)\ ,\quad i,j=1,2,\dots , D-1
\ee
and have zero correlators otherwise.
Consider then the representation
\ba
&& \Psi=-{i\ov 2}(\phi_j \phi_j + D-1)\del\phi_+\
-\phi_j\del\phi_j +\ i \del\phi_-\ ,
\qq
\Phi= -i \del\phi_+\ ,
\nonumber\\
&& P_j= {1\ov \sqrt{2}}(i \del\phi_j - \phi_j \del\phi_+)\ ,
\quad j=1,2,\dots ,D-1\ ,
\label{JHWE1}
\ea
where the presence of the constant $D-1$ in the expression for $\Psi$ is
necessary in order to obtain the correct operator product
algebra in \eqn{oppe1} below.\footnote{In particular, the presence of this
constant is required for the
absence of a second order pole with a $c$-number coefficient
in the operator product expansion of $\Psi$
with itself.}
The stress-energy tensor of the theory with respect to which $\Phi$ and $P_i$
are Virasoro primaries and $\Psi$ is the logarithmic partner of $\Phi$ is
\be
T(z)=\del \phi_+\del\phi_- -\ \ha \del \phi_i \del\phi_i
- {i\ov 2} (D-1) \del^2 \phi_+\
\ee
and the corresponding central charge is $c=D+1$.
Using the free field representation \eqn{JHWE1} we compute the operator
product expansions
\ba
\Psi(z)\Phi(w) & = &  {1\ov (z-w)^2} + {\cal O}(1) \ ,
\nonumber\\
\Psi(z)\Psi(w) & =& (D-1){\ln(z-w)\ov (z-w)^2}
+ 2 \ln(z-w) :\!(P_iP_i)(w)\!:
\nonumber\\
&& \phantom{xxxxxx}
 +{D-1\ov2 } \ln^2(z-w) :\!\Phi^2(w)\!: +\ {\cal O}(1)\ ,
\nonumber\\
\Psi(z) P_i(w)& = & - \ln(z-w) :\!(P_i\Phi)(w)\!: +\ {\cal O}(1) \ ,
\label{oppe1}\\
P_i(z)P_j(w)& = & {\d_{ij}\ov 2 (z-w)^2} + \ha \d_{ij }\ln(z-w)
:\!\Phi^2(w)\!:
+\ {\cal O}(1) \ ,
\nonumber
\ea
which are nothing but the exact operator product expansion algebra
corresponding to the Poisson brackets \eqn{pppoi2}. Notice also
that for $D=2$ this becomes the operator algebra \eqn{oppe2} as
it should be.

\subsection{Correlation functions}

It is possible to compute correlation functions for our theories. The
non-zero two-point
functions are immediately read off from the operator algebra \eqn{oppe1}.
In general, it can be shown that all odd-point functions vanish and in
addition all $2n$-point functions involving more than $n$ insertions of the
field $\Phi$ also vanish.
Computing four-point functions among the logarithmic partners $\Phi$ and $\Psi$
gives (we use below the notation $z_{ij}=z_i-z_j$)
\ba
&& \langle \Phi(z_1)\Phi(z_2)\Phi(z_3)\Psi(z_4)\rangle = 0 \ ,
\nonumber\\
&& \langle \Phi(z_1)\Phi(z_2)\Psi(z_3)\Psi(z_4)\rangle =
{1\ov z_{13}^2 z_{24}^2}+ {1\ov z_{14}^2 z_{23}^2}\ ,
\label{pppp}\\
&&  \langle \Phi(z_1)\Psi(z_2)\Psi(z_3)\Psi(z_4)\rangle = (D-1)\left(
{\ln z_{34}\ov z_{12}^2 z_{34}^2}
+ {\ln z_{24}\ov z_{13}^2 z_{24}^2}
+ {\ln z_{23}\ov z_{14}^2 z_{23}^2}\right)\ ,
\nonumber
\ea
which is in agreement with general expectations for four-point functions
of logarithmic partners in logarithmic conformal
field theories \cite{Flrev}.
We may also show that the four-point function having only the logarithmic
field $\Psi$ appearing, assumes the form given in the literature \cite{Flrev}.
These results agree with those obtained for the $D=2$ case in \cite{BSpp},
which therefore captures the essential characteristics of these correlators.

It is interesting to consider also correlation functions containing the primary
field $P_i$ as well.
A group theory argument shows that if such correlators contain
an odd number of $P_i$'s then they are necessarily zero.
Restricting to four-point functions we find for the (potentially)
non-zero ones
\ba
&& \langle P_i(z_1) P_j(z_2) P_k(z_3) P_l(z_4)\rangle = {1\ov 4}\left(
{\d_{ij}\d_{kl}\ov z_{12}^2 z_{34}^2} +
{\d_{ik}\d_{jl}\ov z_{13}^2 z_{24}^2} +
{\d_{il}\d_{jk}\ov z_{14}^2 z_{23}^2}\right)\ ,
\nonumber\\
&& \langle P_i(z_1) P_j(z_2)\Phi(z_3)\Psi_4(z_4)\rangle =
\ha {\d_{ij}\ov  z_{12}^2 z_{34}^2}\ ,
\label{4pp}\\
&& \langle P_i(z_1) P_j(z_2)\Phi(z_3)\Phi_4(z_4)\rangle =0\ .
\ea
The expression for the four-point function $\langle P_i(z_1) P_j(z_2)\Psi(z_3)
\Psi_4(z_4)\rangle$, which is also non-vanishing, is quite lengthy
so that we give it separately
\ba
&& \langle P_i(z_1) P_j(z_2)\Psi(z_3) \Psi_4(z_4)\rangle = \d_{ij} \times
\nonumber\\
&& \left[{D-1\ov 2} {\ln z_{34}\ov z_{12}^2 z_{34}^2} + \ha  {1\ov
z_{13} z_{14} z_{23} z_{24}}
 + \ha {1\ov z_{13}^2 z_{24}^2}\ln\left(z_{34} z_{12}\ov z_{23} z_{14}\right)
+ \ha {1\ov z_{14}^2 z_{23}^2}\ln\left(z_{34} z_{12}\ov z_{24} z_{13}
\right) \right] \ .
\ea
This expression has the correct behaviour for $z_1\to z_2$ and $z_1\to z_3$
according to the operator algebra \eqn{oppe1}. In particular,
we find that for $z_1\to z_3$ the correlator behaves correctly
as $\ln z_{13}$ and not as apparently looks like, i.e. $1/z_{13}^2 $.

For more general coset models which do not correspond to
symmetric spaces and have, in the notation of \eqn{sttru}, the structure
constants $f_{ij}\neq 0 $ and (or) $f_{ijk}\neq 0$,
constructing the free field representation of the various fields is a bit more
involved and will not be presented here.

\section{Four- and five-dimensional examples}

Let us consider in some more detail the $D+1$
string backgrounds arising from the
$SO(D+1)_N/SO(D)_N\times U(1)_{-N}$ theories at $N\to \infty$.
The lowest dimensional of these models, corresponding to $D=2$,
represents a plane wave background and is the prototype example for the
logarithmic conformal field theories that arose in this limit.
Here we will consider the
cases with $D=3$ and $4$ where the backgrounds corresponding to the coset
non-trivial factor have been worked out in the literature explicitly.
We will see that, the resulting geometries, although they have a null Killing
vector, are not plane waves.

For the case of $SO(4)_N/SO(3)_N \times U(1)_{-N}$ the metric and dilaton
are \cite{Bars1}\footnote{These backgrounds were given for
non-compact versions of these cosets, i.e. $SO(2,2)_{-N}/SO(2,1)_{-N}$ etc.
Analytically continuing to the present compact coset case is trivial.}
\ba
 {1\ov N} ds^2 & = &
-dt^2 + d\th^2 + {\cot^2\th\ov \cos^2\om} d\phi^2
+ \tan^2\th (d\om + \tan\om \cot\phi d\phi)^2  \ ,
\nonumber\\
e^{-2\Phi} & = & e^{-2\Phi_0}\sin^2 2\th \sin^2 \phi \cos^2 \om\ ,
\ea
whereas the antisymmetric tensor is zero.\footnote{This is true for the
general $SO(D+1)_N/SO(N)_N$ coset model.}
We consider the redefinitions
\be
\th={1\ov N} v+u \ ,\qq t=u \ ,\qq \phi={\r\ov \sqrt{N}} \ ,
\qq \om={1\ov \sqrt{N}} {\r_1\ov \r}
\ee
and then take the limit $N\to \infty$. We find the metric and dilaton
\ba
ds^2 & = & 2 dudv + \cot^2u d\r^2 + {\tan^2u \ov \r^2}d\r_1^2 \ ,
\nonumber\\
 e^{-2 \Phi} & = & e^{-2 \Phi_0} \r^2 \sin^2 2 u\ .
\label{jgsdjh}
\ea
where we have also redefined the constant
$e^{2\Phi_0}\to {1\ov N} e^{2\Phi_0}$. Note that this background
has the null Killing vector $\del/\del v$, but it is not a
plane wave since it depends explicitly on a variable other than $u$.
It is worth noticing that it is related via T-duality to the
four-dimensional plane wave background
\ba
ds^2 & = & 2 dudv + \cot^2u (dx_1^2 + dx_2^2)\ ,
\nonumber\\
 e^{-2 \Phi} & = & e^{-2 \Phi_0} \sin^4 u\ .
\label{pplal}
\ea
In order to show that we first perform to the background \eqn{jgsdjh}
a T-duality transformation with
respect to the Killing vector $\del/\del\r_1$. The resulting metric has
a two-dimensional
transverse space of the form $d\r^2+\r^2 d\r_1^2$ which is locally the metric
for $\IR^2$ in polar coordinates.
However, in order to avoid a singularity at $\r=0$ (at constant
$u$) one has
to compactify the variable $\r_1$ to have period $2\pi$. Then, after passing
to a Cartesian coordinate system, we obtain \eqn{pplal}.

For the case of $SO(5)_N/SO(4)_N \times U(1)_{-N}$ the metric and dilaton
are \cite{Bars2}
\ba
 {1\ov N} ds^2 & = & -dt^2 + d\th^2
+\cot^2\th \left({d\phi_1^2\ov \cos^2\om}  + {d\phi_2^2\ov \sin^2\om}\right)
\nonumber\\
&& + \tan^2\th \left(d\om - {\tan\om \sin2\phi_1 d\phi_1 +
\cot\om \sin2\phi_2 d\phi_2 \ov \cos2\phi_1 - \cos2\phi_2}\right)^2\ ,
\label{dd5}\\
e^{-2\Phi} & = & e^{-2\Phi_0}\sin^2 2\th \sin^2\th \sin^2 2\om
(\cos 2\phi_1-\cos 2\phi_2)^2 \ .
\nonumber
\ea
We consider the redefinitions
\be
\th={1\ov N} v+u \ ,\qq t=u \ ,\qq \phi_1={\r\ov \sqrt{N}} \ ,\qq
\phi_2={\r_2\ov N}\ ,
\qq \om=\sqrt{\r_1 - \r_2^2 \ov N \r^2}\
\ee
and then take the limit $N\to \infty$. We find the metric and dilaton
\ba
ds^2 & = & 2 dudv + \cot^2u \left(d\r^2 +{\r^2d\r_2^2\ov \r_1-\r_2^2}\right)
+ \tan^2u {d\r_1^2\ov 4 \r^2(\r_1-\r_2^2)}\ ,
\nonumber\\
 e^{-2 \Phi} & = & e^{-2 \Phi_0} \r^2 (\r_1-\r_2^2)
\sin^2 u\sin^2 2 u\ .
\ea
with zero antisymmetric tensor and where we have redefined the constant
$e^{2\Phi_0}\to {1\ov N^3} e^{2\Phi_0}$. It is obvious that this background is
not related via T-duality to a plane wave.

\section{Conclusions}

We have constructed a novel class of logarithmic conformal field theories
corresponding to strings propagating on backgrounds with a null Killing vector.
These conformal field theories were obtained as a correlated $N\to \infty$
limit of the $G_N/H_N\times U(1)_{-N}$ theories for general groups $G$ and $H$.
We have obtained the
exact operator product expansion algebra of the basic chiral operators,
their representation in terms of free
bosons as well as explicit expressions for the four-point functions.
We also presented the string backgrounds for four- and five-dimensional
examples.
An interesting further
extension would be to replace the $U(1)_{-N}$ factor
by other exact conformal field theories having one time-like variable.

Our logarithmic theories are non-unitary by construction due to the
presence of the time variable. An interesting question is whether
or not imposing the Virasoro constraints in a covariant approach or by
choosing the light-cone gauge we can preserve unitarity in string theory.
In addition, the fact that these logarithmic conformal field theories
correspond
to a large level limit of known coset theories times a free time-like boson,
may provide useful information on how to deal with several technical issues
concerning logarithmic conformal field theories in general. For instance,
how to
correctly combine left and right movers in order to construct local field
theories. We also think that, whatever progress is made towards understanding
issues concerning logarithmic conformal field theories in string
theory, will finally shed further light to the
recent developments on pp-waves and the
AdS/CFT correspondence for highly massive string states
\cite{BMN}, as well as to the occurence of logarithmic behaviour in correlators
of gauge theories related to anomalous dimensions and non-protected operators
(see, for instace, \cite{massimo}).


\vspace{8 mm}

\centerline {\bf Acknowledgments}
\noindent
$\bullet$
This work was supported in part by the European Research and Training Networks
``Superstring Theory" (HPRN-CT-2000-00122) and ``The Quantum Structure of
Space-time" (HPRN-CT-2000-00131). I also acknowledge support by the Greek State
Scholarships Foundation under the contract IKYDA-2001/22,
as well as NATO support
by a Collaborative Linkage Grant under the contract PST.CLG.978785.

\noindent
$\bullet$
This paper is dedicated to the memory of the young colleague Sonia Stanciu
who got involved and contributed at various stages of her carrier
to problems related to plane waves and exact conformal field theories
\cite{FF, SS}.



\begin{thebibliography}{99}
\renewcommand{\baselinestretch}{1}
\normalsize

\bibitem{BSpp}
I. Bakas and K. Sfetsos, {\it PP-waves and logarithmic conformal field
theories},\hfill\break {\tt hep-th/0205006}.

\bibitem{Pen}
R. Penrose, {\it Any space-time has a wave as a limit}, Differential
Geometry and Relativity, Reidel, Dordrecht, 1976.


\bibitem{paraf}
A.B. Zamolodchikov and V.A. Fateev, Sov. Phys. JETP {\bf 62} (1985) 215.

\bibitem{saleur}
L.~Rozansky and H.~Saleur,
Nucl. Phys. {\bf B376} (1992) 461.

\bibitem{Gur}
V. Gurarie, Nucl. Phys. {\bf B410} (1993) 535,
{\tt hep-th/9303160}.

\bibitem{BK}
A. Bilal and I. Kogan, Nucl. Phys. {\bf B449} (1995) 569, {\tt hep-th/9503209}.

\bibitem{Cau}
J.-S. Caux, I. Kogan and A. Tsvelik, Nucl. Phys. {\bf B466}
(1996) 444, {\tt hep-th/9511134}.

\bibitem{KN}
I. Kogan and A. Nichols, JHEP {\bf 0201} (2002) 029,
{\tt hep-th/0112008}.

\bibitem{persis}
A.~M.~Ghezelbash, M.~Khorrami and A.~Aghamohammadi,
Int.J. Mod. Phys. {\bf A14} (1999) 2581,
{\tt hep-th/9807034}.
\bibitem{gardy}
J.L.~Cardy,
J. Phys. {\bf A25} (1992) L201, {\tt hep-th/9111026}.

\bibitem{saleur2}
H.~Saleur,
Nucl. Phys. {\bf B382} (1992) 486,
hep-th/9111007.


\bibitem{Flrev}
M. Flohr, {\it Bits and Pieces in Logarithmic Conformal Field
Theory}, {\tt hep-th/0111228}.

\bibitem{Gilmore}
R. Gilmore, {\it Lie Groups, Lie Algebras and Some of
Their Applications}, John Wiley.

\bibitem{sfe1}
K.~Sfetsos,
Phys. Lett. {\bf B324} (1994) 335,
{\tt hep-th/9311010}.

\bibitem{Olive}
D.I.~Olive, E.~Rabinovici and A.~Schwimmer,
Phys. Lett. {\bf B321} (1994) 361,
{\tt hep-th/9311081}.

\bibitem{sfe2}
K.~Sfetsos,
Phys. Rev. {\bf D50} (1994) 2784,
{\tt hep-th/9402031}.

\bibitem{sfetse}
K.~Sfetsos and A.A.~Tseytlin,
Nucl. Phys. {\bf B427} (1994) 245,
{\tt hep-th/9404063}.


\bibitem{NaWi}
C.R.~Nappi and E.~Witten,
Phys. Rev. Lett. {\bf 71} (1993) 3751,
{\tt hep-th/9310112}.

\bibitem{Mohammedi}
N.~Mohammedi,
Phys. Lett. {\bf B325} (1994) 371,
{\tt hep-th/9312182}.

\bibitem{FF}
J.M.~Figueroa-O'Farrill and S.~Stanciu,
Phys. Lett. {\bf B327} (1994) 40,
{\tt hep-th/9402035}.

\bibitem{kehm}
A. Kehagias and P. Meesen, Phys. Lett. {\bf B331} (1994) 77,
{\tt hep-th/9403041}.

\bibitem{Forgacs}
P.~Forgacs, P.A.~Horvathy, Z.~Horvath and L.~Palla,
Heavy Ion Phys. {\bf 1} (1995) 65,
{\tt hep-th/9503222}.

\bibitem{Antoniadis}
I.~Antoniadis and N.~A.~Obers,
Nucl. Phys. {\bf B423} (1994) 639,
{\tt hep-th/9403191}.

\bibitem{claspara}
K.~Bardacki, M.J.~Crescimanno and E.~Rabinovici,
Nucl. Phys. {\bf B344} (1990) 344.

\bibitem{kill1}
R. G\"uven, Phys. Lett. {\bf B191} (1987) 275.

\bibitem{kill0}
D. Amati and C. Klimcik, Phys. Lett. {\bf B219} (1989) 443; \hfill\break
G. Horowitz and A. Steif,
Phys. Lett. {\bf 64} (1990) 260 and Phys. Rev. {\bf D42}
(1990) 1950.

\bibitem{kill01}
H.J. de Vega and N. Sanchez, Nucl. Phys. {\bf B317} (1989) 706 and
Phys. Rev. Lett. {\bf 65} (1990) 1517.

\bibitem{kill02}
J. Garriga and E. Verdaguer, Phys. Rev. {\bf D43} (1991) 391;
R.E. Rudd, Nucl. Phys. {\bf B352} (1991) 489
and 
Nucl. Phys. {\bf B427} (1994) 81, {\tt hep-th/9402106};\hfill\break
C. Duval, G.W. Gibbons and P.A. Horvathy,
Phys. Rev. {\bf D43} (1991) 3907;\hfill\break
C. Duval, Z. Horvathy and P.A. Horvathy,
Phys. Lett. {\bf B313} (1993) 10.

\bibitem{kill03}
A.A. Tseytlin, Nucl. Phys. {\bf B390} (1993) 153, {\tt hep-th/9209023}
and Phys. Rev. {\bf D47} (1993) 3421, {\tt hep-th/9211061}.

\bibitem{kill2}
E.A. Bergshoef, R. Kallosh and T. Ortin, Phys. Rev. {\bf D47} (1993) 5444,
\hfill\break {\tt hep-th/9212030}.


\bibitem{karabali}
D.~Karabali, Q.H.~Park, H.J.~Schnitzer and Z.~Yang,
Phys. Lett. {\bf B216} (1989) 307;\hfill\break
D.~Karabali and H.J.~Schnitzer,
Nucl. Phys. {\bf B329} (1990) 649.


\bibitem{parfree}
A. Gerasimov, A. Marshakov and A. Morozov, Nucl. Phys. {\bf B328} (1989)
664; \hfill\break
D. Nemeschansky, Phys. Lett. {\bf B224} (1989) 121; \hfill\break
J. Distler and Z. Qiu, Nucl. Phys. {\bf B336} (1990) 533; \hfill\break
O. Hern\`andez, Phys. Lett. {\bf B233} (1989) 355.


\bibitem{Bars1}
I.~Bars and K.~Sfetsos,
Mod. Phys. Lett. {\bf A7} (1992) 1091,
{\tt hep-th/9110054}.


\bibitem{Bars2}
I.~Bars and K.~Sfetsos,
Phys. Lett. {\bf B277} (1992) 269,
{\tt hep-th/9111040}.


\bibitem{KiKouLu}
E.~Kiritsis and C.~Kounnas,
Phys. Lett. {\bf B320} (1994) 264
[Addendum-ibid. {\bf B325} (1994) 536],
{\tt hep-th/9310202};\hfill\break
E.~Kiritsis, C.~Kounnas and D.~Lust,
Phys. Lett. {\bf B331} (1994) 321,
{\tt hep-th/9404114}.


\bibitem{BMN}
D.~Berenstein, J.~Maldacena and H.~Nastase,
JHEP {\bf 0204} (2002) 013,\hfill\break
{\tt hep-th/0202021}.

\bibitem{massimo}
M.~Bianchi, S.~Kovacs, G.~Rossi and Y.S.~Stanev,
JHEP {\bf 9908} (1999) 020,
{\tt hep-th/9906188};
M.~Bianchi, B.~Eden, G.~Rossi and Y.~S.~Stanev,
{\it On operator mixing in $N = 4$ SYM},
{\tt hep-th/0205321};
G.~Arutyunov, S.~Penati, A.C.~Petkou, A.~Santambrogio and E.~Sokatchev,
{\it Non-protected operators in $N=4$ SYM and multiparticle states of
$AdS_5$  SUGRA}, {\tt hep-th/0206020}.


\bibitem{SS}
J.M.~Figueroa-O'Farrill and S.~Stanciu,
J. Math. Phys. {\bf 37} (1996) 4121,\hfill\break {\tt hep-th/9506152}
and
Nucl. Phys. {\bf B458} (1996) 137, {\tt hep-th/9506151};\hfill\break
S.~Stanciu and A.~A.~Tseytlin,
JHEP {\bf 9806} (1998) 010, {\tt hep-th/9805006};\hfill\break
J.M.~Figueroa-O'Farrill and S.~Stanciu,
JHEP {\bf 0001} (2000) 024, {\tt hep-th/9909164}.


\end{thebibliography}
\end{document}